\documentclass[twocolumn,pra,superscriptaddress,showpacs,epsf]{revtex4}
\usepackage{graphicx}
\usepackage{epsf}
\usepackage{amsmath}
\usepackage[dvips]{color}
\usepackage{pifont}
\input{colordvi.sty}

\begin{document}
\definecolor{blue}{rgb}{0.15,0.1,0.7}
\definecolor{red}{rgb}{0.7,0.1,0.15}
\definecolor{green}{rgb}{0.15,0.7,0.15}
\newcommand{\blue}[1]{{\textcolor{blue}{ \it\small #1}}}
\newcommand{\add}[1]{{\textcolor{red}{ \it\small #1}}}
\newcommand{\Q}{{\textcolor{red}{???? }}}
\title{Cooling a mechanical resonator by
quantum interference in a triple quantum dot}
\date{\today}
\author{Shi-Hua Ouyang}
\affiliation{Department of Applied Physics, Hong Kong Polytechnic
University, Hung Hom, Hong Kong, China} \affiliation{Department of
Physics and Surface Physics Laboratory (National Key Laboratory),
Fudan University, Shanghai 200433, China}
\author{Chi-Hang Lam}
\affiliation{Department of Applied Physics, Hong Kong Polytechnic
University, Hung Hom, Hong Kong, China}
\author{J. Q. You}
\affiliation{Department of Physics and Surface Physics Laboratory
(National Key Laboratory), Fudan University, Shanghai 200433, China}

\begin{abstract}

We propose an approach to cool a mechanical resonator (MR) via
quantum interference in a triple quantum dot (TQD) capacitively
coupled to the MR. The TQD connected to three electrodes is an
electronic analog of a three-level atom in $\Lambda$ configuration.
The electrons can tunnel from the left electrode into one of the two
dots with lower-energy states, but can only tunnel out from the
higher-energy state at the third dot to the right electrode. When
the two lower-energy states are tuned to be degenerate, an electron
in the TQD can be trapped in a superposition of the degenerate
states called the dark state. This effect is caused by the
destructive quantum interference between tunneling from the two
lower-energy states to the higher-energy state. Under this
condition, an electron in the dark state readily absorbs an energy
quantum from the MR. Repeating this process, the MR can be cooled to
its ground state. Moreover, we propose a scheme for verifying the
cooling result by measuring the current spectrum of a charge
detector adjacent to a double quantum dot coupled to the MR.
\end{abstract}

\maketitle

\section{Introduction}
Mechanical resonators (MRs) with a high resonant frequency and a
small mass have wide applications and are attracting considerable
recent attentions~\cite{Huang03,Schwab05}. Technically, these MRs
can be used as ultrasensitive sensors in high-precision displacement
measurements~\cite{LaHaye04}, detection of gravitational
waves~\cite{BraginskyPLA02} or mass detection~\cite{Buks06}. Also,
quantized MRs can be useful in quantum information processing.
Indeed, quantized motion of buckling nanoscale bars has been
proposed for qubit implementation~\cite{SavelNJP,Savel07} and also
for creating quantum entanglement~\cite{Vitali07,Hartmann08,Tian04}.
However, for all these applications, a basic prerequirement is that
the dynamics of the MRs must approach the quantum regime.

Quantum behaviors of a MR are usually suppressed by the coupling to
its environment. One way to approach the quantum regime is to
increase its resonant frequency so that an energy quantum of the MR
becomes larger than the thermal energy. Recently, MRs based on
metallic beams~\cite{TF08} and carbon nanotubes~\cite{Huttel09} have
been developed, which have resonance frequencies of several hundred
megahertz. However, for a MR with a frequency of $200$~MHz, a
temperature lower than $10$~mK (below the present dilution
refrigerator temperature) is required to maintain the MR at the
quantum regime. To attain the quantum regime, one needs to cool the
MR further via coupling to an optical or an on-chip electronic
system. Numerous experiments on cooling a single MR via radiation
pressure
or dynamical backaction have been reported (see, e.g.,~\cite{Metzger04,Gigan06,%
Kleckner06,Arcizet06,Naik06,Schliesser06,Poggio07,Lehnert08,Kippenberg05}).
Theoretically, cooling by coupling to a
Cooper pair box~\cite{Zhang05} or to a three-level flux
qubit~\cite{You08} via periodic resonant coupling have also been proposed. In these schemes,
a strong resonant coupling between the MR and the qubit is required
to cool the MR to its ground state.

\subsection{Sideband cooling of a MR}
In the weak coupling regime, a conventional method for cooling the
MR is the sideband cooling approach (see, e.g.,
Ref.~\cite{Wilson04,Wilson07,Marquardt07,ASchliesser08,Yong08,Ouyang09,Zippilli09,Grajcar08}).
In this case, a MR is coupled to a two-level system (TLS) in which
the two states
can be electronic states in quantum
dots~\cite{Wilson04,Ouyang09,Zippilli09}, photonic states in a
cavity~\cite{Wilson07,Marquardt07,ASchliesser08,Yong08}, or charge
states in superconducting qubits~\cite{You05}. In order to achieve
ground-state cooling in the sideband cooling approach, the
resolved-sideband cooling condition $\omega_m\gg\Gamma$ (with
$\omega_m$ denoting the oscillating frequency of the MR and $\Gamma$
the decay rate of the TLS) must be followed in order to selectively
drive the lowest sideband of the TLS.
Then, the excitation of the TLS from the ground state $|g\rangle$ to
the excited state $|e\rangle$ and the subsequent decay from this
excited state to the ground state will, on average, decrease an
energy quantum in the MR. This process can be described by
$|g,n\rangle\rightarrow|e,n-1\rangle\rightarrow|g,n-1\rangle$, where
$n$ denotes to the state with $n$ phonons. However, the frequency of
a typical MR is about $100$~MHz \cite{TF08,Huttel09}. It is in
general of the same order of the decay rate of the two-level system.
This indicates that the resolved-sideband cooling condition is not
easy to fulfill. Violating the condition means that the
processes of a carrier transition and a subsequent sideband
transition (i.e.,
$|g,n\rangle\rightarrow|e,n\rangle\rightarrow|g,n+1\rangle$) will
occur. This will heat up the MR instead and suppress any ground-state
cooling~\cite{ASchliesser08}.

\subsection{Cooling atomic motion via quantum interference in a three-level atom}

For laser cooling of atoms, an alternative approach~\cite{Morigi00}
based on quantum interference in the internal degrees of freedom of
the atoms without the need to follow the resolved-sideband cooling
condition has been proposed. In this approach, an additional state
is coupled to the excited state $|3\rangle$ of the TLS to form a
$\Lambda$-shaped three-level system [see Fig.~\ref{fig1}(b)]. The
two lower-energy states in this three-level system are tuned to be
degenerate. Due to dissipation of the excited state, the atom will
eventually arrive at a particular superposition of the two
lower-energy states which is orthogonal to the excited state.  This
phenomenon results from the destructive quantum interference between
the two transitions $|1\rangle\rightarrow|3\rangle$ and
$|2\rangle\rightarrow|3\rangle$ and the superposition state is
called the dark state~\cite{Scully}. When atomic motion is also
considered, the carrier transition
($|1,n\rangle\rightarrow|3,n\rangle$) and thus the heating process
of the atomic motion in the sideband cooling scheme is hence
suppressed. It was shown that atomic motion can be cooled to its
ground state in the non-resolved sideband regime~\cite{Morigi00}.

\subsection{Cooling a MR via quantum interference in a triple quantum dot}

\begin{figure}
\includegraphics[width=2.0in,
bbllx=138,bblly=218,bburx=448,bbury=618]{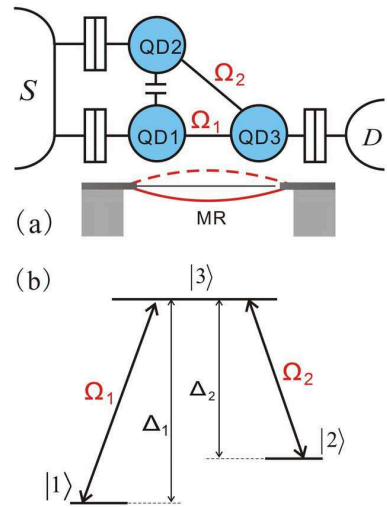} \caption{(color
online)~Schematic diagram of a TQD system. Dots $1$ and $2$ are both
tunnel-coupled to dot $3$ (with interdot coupling strengths
$\Omega_1$ and $\Omega_2$, respectively) while they are only
capacitively coupled to each other. A MR is capacitively coupled to
dots $1$ and $3$ of the TQD. (b) A three-level system in $\Lambda$
configuration. The energy detunings between the two lower-energy
states and the third excited state are respectively $\Delta_1$ and
$\Delta_2$. The coupling strength between the state $|1\rangle$
($|2\rangle$) and the state $|3\rangle$ is $\Omega_1$ ($\Omega_2$).
\label{fig1}}
\end{figure}

In the present work, we propose a new scheme to cool a MR via
capacitive coupling to a triple quantum dot (TQD) schematically
displayed in Fig.~\ref{fig1}(a). We consider the strong
Coulomb-blockade regime so that at most one electron is allowed to
present at one time in the TQD. The TQD acts as a three-level system
in $\Lambda$ configuration, in which the two dot states $|1\rangle$
and $|2\rangle$ (i.e., the single-electron orbital states in dots 1
and 2) are coupled to a third (excited) state $|3\rangle$ via two
tunnel barriers [see Fig.~\ref{fig1}(a)]. Here, the degrees of
freedom of the MR is analogous to the motional degrees of freedom of
the atoms discussed above and the TQD is an electronic analog of a
three-level atom. We will show that by properly tuning the gate
voltages, one can degenerate the two lower-energy states and obtain
a dark state in the TQD. By capacitively coupling to the TQD, the MR
can be cooled to its ground state, in full analogy to
the slowing down of the atoms via quantum interference. Comparing
with the cooling of atoms~\cite{Morigi00}, our approach has the
following potential advantages:~(i)~Our cooling system is completely
electronic and can be conveniently fabricated on a chip. (ii)~Simply
by adjusting the gate voltages, it is easy to achieve the two
degenerate lower-energy states required for realizing destructive
quantum interference in the TQD. Moreover, in contrast to the cooling
of a
MR by coupling it to a superconducting qubit~\cite{Xia09}, the decay
rate $\Gamma$ of the higher-energy state of the TQD, which is equal
to the rate of electrons tunneling from the TQD to the electrode, is
tunable in our case by varying the gate voltage.

Moreover, we also propose a method to verify whether the MR is
successfully cooled by coupling it to a double quantum dot (DQD).
This DQD can be reduced from the TQD by applying appropriate gate
voltages. When the DQD and the MR is tuned into a strongly
dispersive regime in which the transition frequency difference
between the two subsystems is much larger than the coupling strength
between them, the coupling between the MR and the DQD only yields a
phonon-number-dependent Stark shift to the transition frequency of
the DQD. This Stark shift corresponds to the shift of the resonant
peak in the current spectrum of a charge detector. Thus, by
measuring the shift of the resonant peak, one can readout the
phonon-number state and examine whether the MR is successfully
cooled or not.

This paper is organized as follows. Section II introduces a
microscopic model for the coupled MR-TQD system. We show that the
TQD is an electronic analog of a three-level atom driven by two
electromagnetic fields. Also, we show how the TQD evolves
into the dark state. In Sec. III, we derive a master equation to
describe the quantum dynamics of the coupled MR-TQD system. With
this master equation, we further derive in Sec.~IV the master
equation of the MR by eliminating the TQD degrees of freedom.
Moreover, we calculate the steady-state average phonon occupancy of
the MR and show that the MR can indeed be cooled to its ground state
by using the quantum interference in the TQD. In Sec.~V, we propose
a method to verify if the MR is successfully cooled by measuring the
full-frequency current spectrum of a charge detector. Section VI
summarizes our results. In the appendix, we give a detailed
derivation of the master equation for the reduced density matrix of
the MR.

\section{A mechanical resonator coupled to a triple quantum dot}

\subsection{Model}

The device layout of a MR coupled to a TQD is shown in
Fig.~\ref{fig1}(a). The TQD is connected to three electrodes via
tunneling barriers. In the TQD, dots $1$ and $2$ are only
capacitively coupled to each other and electrons cannot tunnel
directly between them. Such capacitively coupled dots have already
been achieved in experiments (see, e.g., \cite{Clure07}). In
contrast, electrons can tunnel between dots $1$ and $3$ as well as
between dots $2$ and $3$. Here we focus on the strong
Coulomb-blockade regime, so that at most a single electron is
allowed in the TQD. Thus, only four electronic states need to be
considered in the TQD, i.e., the vacuum state $|0\rangle$, and
states $|1\rangle$, $|2\rangle$ and $|3\rangle$ corresponding to a
single electron in the respective dot.
The
MR is capacitively coupled to dots $1$ and $3$ and this is schematically
shown in Fig.~\ref{fig1}(a).

The total Hamiltonian of the whole system reads
\begin{equation}
H_{\rm{total}}\!=\!H_0+H_{\rm{int}}+H_{\rm{T}}+H_{\rm ep}.
\end{equation}
The unperturbed Hamiltonian $H_0$ is defined by
\begin{equation}
H_0\!=\!H_{\rm{leads}}+H_{\rm{TQD}}+H_{\rm{R}}+H_{\rm ph},
\label{H0}
\end{equation}
where terms on the R.H.S. of Eq.~(\ref{H0}) denote Hamiltonians of
the electrodes, the TQD, the MR and the thermal bath given by
\begin{eqnarray}
H_{\rm{leads}}\!&=\!&\sum_{i k}E_{i k}c_{i k}^{\dagger}%
c_{i k},\\
H_{\rm{TQD}}\!\!&\!=\!&\!\!-\Delta_1a_1^\dagger a_1\!-\!\Delta_2a_2^\dagger a_2
\!+\!(\Omega_1a_1^\dagger a_3+\Omega_2a_2^\dagger a_3+{\rm H.c.}),\label{H-TQD}
\nonumber\\
&&\\
H_{\rm{R}}\!&=\!&\omega_mb^{\dagger}b,\label{MR}
\\
H_{\rm ph}\!&=\!&\sum_{q}\omega_qb_q^\dagger b_q.\label{thermalbath}
\end{eqnarray}
We have put $\hbar=1$ and the energy of the state $|3\rangle$ is
chosen as the zero-energy point. $c_{i k}^{\dagger}$ ($c_{i k}$) is
the creation (annihilation) operator of an electron with momentum
$k$ in the $i$th electrode ($i=1,2$ or $3$) and $a_i^{\dagger}$
creates an electron in the $i$th dot. The phonon operators
$b^\dagger$ and $b$ respectively create and annihilate an excitation
of frequency $\omega_m$ in the MR. In Eq.~(\ref{thermalbath}), the
thermal bath is modeled as a bosonic bath with $b_q^\dagger$ ($b_q$)
being the creation (annihilation) operator at freqency $\omega_q$.

The electromechanical coupling between the MR and dots 1 and 3 of the TQD is given by
\begin{eqnarray}
H_{\rm{int}}=-g\;(a_3^\dagger a_3-a_1^\dagger a_1)(b^{\dagger}+b),\label{coupling}
\end{eqnarray}
%
with a coupling strength $g=\eta\omega_m$. For a typical electromechanical
coupling, $\eta$ $\sim 0.1$ (see, e.g., Ref.~\cite{Neill09}). The
tunneling coupling between the TQD and the electrodes is
\begin{eqnarray}
H_{\rm{T}}=\sum_{ik}(\Omega_{ik}\;a_{i}^{\dagger}\,c_{ik}+\rm{H.c.}),
\end{eqnarray}
where $\Omega_{ik}$ characterizes the coupling strength between the
$i$th dot and the associated electrode via tunneling barrier.
Moreover, the coupling of the MR to the outside thermal bath is
characterized by
\begin{equation}
H_{\rm ep}=\sum_q\Omega_q(b_q^\dagger b+{\rm H.c.}),
\end{equation}
with a coupling strength $\Omega_q$.

\subsection{Analogy between TQD and $\Lambda$-type three-level atom  in two driving electromagnetical fields}

We now show that in the absence of the MR, our TQD system is
analogous to a typical $\Lambda$-type three-level atom in the
presence of two classical electromagnetical fields. This
field-driven three-level system is often used in quantum optics for
producing a dark state (see, e.g., \cite{Scully}). The Hamiltonian
of the field-driven $\Lambda$-type three-level system can be written
as
\begin{eqnarray}
H_{\Lambda}\!&\!=\!&\!\omega_1a_1^\dagger a_1+\omega_2a_2^\dagger a_2+\omega_3a_3^\dagger a_3
\nonumber\\&&
+\Omega_a\cos(\omega_at)(a_1^\dagger a_3+a_3^\dagger a_1)
\nonumber\\&&
+\Omega_b\cos(\omega_bt)(a_2^\dagger a_3+a_3^\dagger a_2),\label{H-lambda}
\end{eqnarray}
where $\omega_i$ ($i=1,2$ or $3$) is the energy of the $i$th
state in the three-level system. Also, $\omega_a$ and $\omega_b$ are the
frequencies of the two driving fields and $\Omega_a$ and $\Omega_b$
are the corresponding driving strengths. In order to eliminate the
time-dependence of the Hamiltonian in Eq. (\ref{H-lambda}), we
transform the system into a rotating frame defined by $U_R=e^{iRt}$
with
\begin{equation}
R=\omega_aa_1^\dagger a_1+\omega_ba_2^\dagger
a_2-\omega_3(a_1^\dagger a_1+a_2^\dagger a_2+a_3^\dagger a_3).
\end{equation}
The transformed Hamiltonian is
\begin{equation}
\widetilde{H}_\Lambda=U_R^{-1}H_\Lambda U_R+i\dot{U}_R^{-1}U_R,\label{H-UR}
\end{equation}
where the first term is evaluated as (within the rotating-wave
approximation)
\begin{eqnarray}
U_R^{-1}H_\Lambda U_R\!&\!=\!&\!\omega_1a_1^\dagger a_1+\omega_2a_2^\dagger a_2+\omega_3a_3^\dagger a_3
\nonumber\\&&
+\frac{\Omega_a}{2}(a_1^\dagger a_3+a_3^\dagger a_1)
+\frac{\Omega_b}{2}(a_2^\dagger a_3+a_3^\dagger a_2),
\nonumber\\&&
\end{eqnarray}
and the second term gives
\begin{eqnarray}
i\dot{U}_R^{-1}U_R\!&=&\!\omega_aa_1^\dagger a_1+\omega_ba_2^\dagger
a_2-\omega_3(a_1^\dagger a_1+a_2^\dagger a_2+a_3^\dagger a_3).
\nonumber\\&&
\end{eqnarray}
Thus,
Eq. (\ref{H-UR}) reduces to
\begin{eqnarray}
\widetilde{H}_\Lambda\!&=&\!-\Delta_1a_1^\dagger a_1-\Delta_2a_2^\dagger a_2
\nonumber\\&&
+\Omega_1(a_1^\dagger a_3+a_3^\dagger a_1)
+\Omega_2(a_2^\dagger a_3+a_3^\dagger a_2),\label{H-RWA}
\end{eqnarray}
where $\Delta_1=\omega_3-\omega_1-\omega_a$ and
$\Delta_2=\omega_3-\omega_2-\omega_b$ are the frequency detunings
while $\Omega_1=\Omega_a/2$ and $\Omega_2=\Omega_{b}/2$ are the
effective driving strengths of the two fields. It is now  clear that
the Hamiltonian in Eq. (\ref{H-RWA}) is formally equivalent to that
of the TQD in Eq.~(\ref{H-TQD}). This shows that the TQD we propose
here is an electronic analog of a $\Lambda$-type three-level atom
driven by two electromagnetic fields.

\subsection{``Dark state'' in the TQD}

From the study of quantum optics, the existence of a dark state
in a $\Lambda$-type three-level atom
when the
lower-energy states become degenerate, i.e., $\Delta_1=\Delta_2$,
is able to suppress absorption or emission.
Below we demonstrate that a similar dark state also exists in the
TQD~\cite{Michaelis06}.

\begin{figure}
\includegraphics[width=2.50in,
bbllx=302,bblly=208,bburx=538,bbury=384]{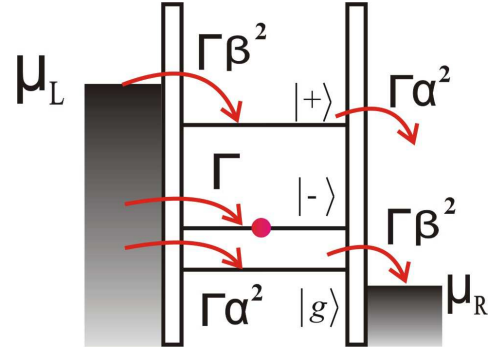} \caption{~(color
online) Effective tunneling processes of electrons through a TQD
represented in the eigenstate basis. An electron can tunnel from the
left electrode into the three eigenstates $|+\rangle$, $|-\rangle$
and $|g\rangle$, with rates $\Gamma\beta^2$, $\Gamma$ and
$\Gamma\alpha^2$, respectively. Note that the total tunneling rate
is $2\Gamma$ because an electron tunnels from the left electrode to
the TQD via two tunnel barriers (each having a tunneling rate
$\Gamma$). In the eigenstate $|+\rangle$ ($|g\rangle$), it will
tunnel out to the right electrode with a rate $\Gamma\alpha^2$
($\Gamma\beta^2$). However, if the electron occupies the dark state
$|-\rangle$, no further tunneling occurs and the electron is
trapped.} \label{fig2}
\end{figure}

After tracing over the degrees of freedom of the electrodes, the
quantum dynamics of the TQD in the absence of the MR is described by
\begin{eqnarray}
\dot{\rho}_d\!&\!=\!&\!\mathcal{L}_{\rm TQD}\rho_d
\nonumber\\
\!&\!=\!&\!-i[H_{\rm TQD},\rho_d]+\Gamma_1\mathcal{D}[a_1^\dagger]\rho
+\Gamma_2\mathcal{D}[a_2^\dagger]\rho_d+\Gamma_3\mathcal{D}[a_3]\rho_d,
\nonumber\\
\label{ME-TQD}
\end{eqnarray}
where $\rho_d$ is the reduced density matrix of the TQD and
$\Gamma_i$ ($i=1,~2$ or $3$) is the rate for electrons tunneling
into or out of the $i$th dot. The notation $\mathcal{D}$ for any
operator $A$ is given by
\begin{equation}
\mathcal{D}[A]\rho=A\rho
A^{\dagger}-\frac{1}{2}[A^{\dagger}A\rho+\rho A^{\dagger}A].
\end{equation}
Considering equal energy detunings of the lower energy states $|1\rangle$ and
$|2\rangle$ with respect to the excited state $|3\rangle$,
i.e., $\Delta_1=\Delta_2=\Delta$, the eigenstates $|g\rangle$,
$|-\rangle$ and $|+\rangle$ of the TQD become
\begin{eqnarray}
\label{eigenstates}
|g\rangle&\!=&\!\beta|3\rangle-\frac{\alpha}{\Omega}(\Omega_1|1\rangle+\Omega_2|2\rangle),
\nonumber\\
|-\rangle&\!=&\!\frac{1}{\Omega}(\Omega_2|1\rangle-\Omega_1|2\rangle),
\nonumber\\
|+\rangle&\!=&\!\alpha|3\rangle+\frac{\beta}{\Omega}(\Omega_1|1\rangle+\Omega_2|2\rangle),
\end{eqnarray}
where $\alpha=\cos(\theta/2)$, $\beta=\sin(\theta/2)$,
$\tan\theta=2\Omega/\Delta$, and
$\Omega=\sqrt{\Omega_1^2+\Omega_2^2}$. The corresponding
eigenenergies are
\begin{eqnarray}
E_{\rm g}=-\frac{\Delta+\phi}{2},~~E_-=-\Delta,~~E_+=-\frac{\Delta-\phi}{2},
\end{eqnarray}
with $\phi=\sqrt{\Delta^2+4\Omega^2}$. For simplicity, we consider
equal couplings of the three quantum dots to the corresponding
electrodes, i.e., $\Gamma_1=\Gamma_2=\Gamma_3\equiv\Gamma$. Based on
the eigenstate basis of the TQD in Eq. (\ref{eigenstates}), the
equations of motion for the reduced density matrix elements of the
TQD are obtained from Eq.~(\ref{ME-TQD}) as
\begin{eqnarray}
\dot{\rho}_{00}\!&\!=\!&\!-2\Gamma\rho_{00}+\Gamma\beta^2\rho_{gg}
+\Gamma\alpha^2\rho_{++}+\Gamma\alpha\beta(\rho_{+g}+\rho_{g+}),
\nonumber\\
\dot{\rho}_{gg}\!&\!=\!&\!\Gamma\alpha^2\rho_{00}-\Gamma\beta^2\rho_{gg}
-\frac{\Gamma}{2}\alpha\beta(\rho_{+g}+\rho_{g+}),
\nonumber\\
\dot{\rho}_{--}\!&\!=\!&\!\Gamma\rho_{00},
\nonumber\\
\dot{\rho}_{++}\!&\!=\!&\!\Gamma\beta^2\rho_{00}-\Gamma\alpha^2\rho_{++}
-\frac{\Gamma}{2}\alpha\beta(\rho_{+g}+\rho_{g+}),
\nonumber\\
\dot{\rho}_{+g}\!&\!=\!&\!-i(E_{\rm +}-E_{\rm g})\rho_{+g}-\frac{\Gamma}{2}\rho_{+g}
\nonumber\\
&&-\frac{\Gamma}{2}\alpha\beta(\rho_{++}+\rho_{gg})
-\Gamma\alpha\beta\rho_{00}.\label{EOM-eigenstate}
\end{eqnarray}
Figure~\ref{fig2} schematically show effective electron tunneling
processes through the TQD as described by
Eq.~(\ref{EOM-eigenstate}).
Starting from an initially empty TQD, an electron can tunnel from
the left electrode into any of the three eigenstates, with tunneling
rates $\Gamma\beta^2$, $\Gamma$ and $\Gamma\alpha^2$ for eigenstates
$|+\rangle$, $|-\rangle$ and $|g\rangle$, respectively. An electron
in the eigenstate $|+\rangle$ ($|g\rangle$) then tunnels out of the
TQD to the right electrode with a rate $\Gamma\alpha^2$
($\Gamma\beta^2$). However, if the electron occupies the state
$|-\rangle$, no further tunneling occurs because, being orthogonal
to $|3\rangle$, it is decoupled from the right electrode. Therefore,
an electron in the TQD will be trapped in the state $|-\rangle$,
which is called the dark state in quantum optics~\cite{Michaelis06}.
This dark state results from the destructive quantum interference
between the transition $|1\rangle\rightarrow|3\rangle$ (i.e., the
electron tunneling from state $|1\rangle$ to state $|3\rangle$ in
the TQD system) and the transition $|2\rangle\rightarrow|3\rangle$
(i.e., the electron tunneling from state $|2\rangle$ to state
$|3\rangle$).
%

\section{Effective Hamiltonian and master equation for the coupled
MR-TQD system}

We now study the coupled MR-TQD system. Rather than analyzing
directly the energy exchange between the MR and the TQD which
involves tedious algebra, we apply a canonical transform $U=e^S$ on
the whole system, where
\begin{equation}
S=\eta(a_3^\dagger a_3-a_1^\dagger a_1)(b-b^\dagger).
\end{equation}
The transformed Hamiltonian is given by
\begin{eqnarray}
H\!&\!=\!&\!UH_{\rm{total}}U^{\dagger} \nonumber\\
\!&\!=\!&\!H_{\rm leads}+H_{\rm ph}+H_{\rm ep}
+\omega_mb^{\dagger}b
-\Delta_1a_1^\dagger a_1-\Delta_2a_2^\dagger a_2
\nonumber\\
&&
+[\Omega_1a_1^\dagger a_3B^2+\Omega_2a_2^\dagger a_3B+{\rm H.c.}],
\nonumber\\
&&+\sum_{k}\big[\Omega_{1k}\;a_{1}^{\dagger}\,c_{1k}B+\Omega_{2k}\;a_{2}^{\dagger}\,c_{2k}\nonumber\\
&&+\Omega_{3k}\;a_{3}^{\dagger}\,c_{3k}B^\dagger+\rm{H.c.}\big],\label{H-Us}
\end{eqnarray}
where we have neglected a small level shift of $\eta^2\omega_m$ to
the states $|1\rangle$ and $|3\rangle$ and we have also defined
\begin{equation}
B=\exp[{-\eta(b-b^{\dagger})}].\label{B}
\end{equation}
To describe the quantum dynamics of the coupled MR-TQD system, we
derive a master equation (under the Born-Markov
approximation) by tracing over the degrees of freedom of both the
electrodes and the thermal bath. Up to second order in $\eta$,
the master equation can be written as
\begin{eqnarray}
\frac{d\rho}{dt}\!&=&\!-i\omega_m[b^\dagger b,\rho]-i[H_{\rm TQD},\rho]
\nonumber\\&&
-i[V(b^\dagger-b),\rho]+\mathcal{L}_{\rm{T}}\rho+\mathcal{L}_{\rm{D}}\rho,\label{ME}
\end{eqnarray}
where
\begin{eqnarray}
V\!&=\!&\eta\big[2\Omega_1(a_1^\dagger a_3-a_3^\dagger a_1)
+\Omega_2(a_2^\dagger a_3-a_3^\dagger a_2)\big],
\nonumber\\&&
\\
\mathcal{L}_{\rm T}\rho\!&=\!&\Gamma\mathcal{D}[a_1^\dagger]\rho+\Gamma\mathcal{D}[a_2^\dagger]\rho
+\Gamma\mathcal{D}[a_3]\rho
\nonumber\\
&&+\;\eta^2\Gamma_1\big(\mathcal{D}[a_1^\dagger b^\dagger]\rho+\mathcal{D}[a_1^\dagger b]\rho\big)
\nonumber\\
&&+\;\eta^2\Gamma_3\big(\mathcal{D}[a_3 b^\dagger]\rho+\mathcal{D}[a_3 b]\rho\big),
\\
\mathcal{L}_{\rm{D}}\rho&=&\gamma[n(\omega_m)+1]\mathcal{D}[b]\rho
+\gamma n(\omega_m)\mathcal{D}[b^\dagger]\rho.\label{Ld}
\end{eqnarray}
Here, the Liouvillian operator $\mathcal{L}_{\rm{T}}\rho$ accounts for
the dissipation due to the electrodes and $\mathcal{L}_{\rm{D}}\rho$
represents the dissipation at the MR induced by the thermal bath.
Also, $\gamma$ denotes the decay rate of excitations in the MR induced by the
thermal bath and $n(\omega_m)$ is the average boson number at
frequency $\omega_m$ in the thermal bath.

\section{Ground-state cooling of the MR}

\subsection{Master equation for the reduced density matrix of the MR}

In the limit $\gamma\ll g\ll\omega_m$, the TQD is weakly coupled to the MR
and can be regarded as part of the environment experienced by the MR.
The degrees of freedom of the TQD can then be adiabatically
eliminated~\cite{Wilson07,Cirac92} and the master equation for the
reduced density matrix $\mu$ of the MR is given by (see Appendix)
\begin{eqnarray}
\dot{\mu}\!&\!=\!&\!-i(\omega_m+\delta_m)[b^{\dagger}b,\,\rho]+\frac{1}{2}%
\left\{\gamma[n(\omega_m)+1]+A_-(\omega_m)\right\}
\nonumber\\
&&\times[2b\mu b^{\dagger}-(b^{\dagger}b\mu+\mu b^{\dagger}b)]\nonumber\\
&&+\frac{1}{2}[\gamma n(\omega_m)+A_+(\omega_m)][2b^{\dagger}\mu
b-(bb^{\dagger}\mu+\mu bb^{\dagger})],\label{ME-MR}
\end{eqnarray}
where $\delta_m$ is the driving-induced shift of the MR frequency.
In Eq.~(\ref{ME-MR}), the additional terms $A_+$ and $A_-$ are
induced by the coupling with the TQD. With this master equation, one
obtains the equation of motion for the phonon-number-probability
distribution, $p_n=\langle n|\mu|n\rangle$, of the MR:
\begin{eqnarray}
\frac{dp_n}{dt}&=&\big\{\gamma[n(\omega_m)+1]+A_-\big\}[(n+1)p_{n+1}-np_n]
\nonumber\\
&&+[\gamma n(\omega_m)+A_+][np_{n-1}-(n+1)p_n],\label{pn}
\end{eqnarray}
Moreover, the equation of motion for the average phonon number,
$\langle n\rangle=\sum_nnp_n$, in the MR can be obtained from
Eq.~(\ref{pn}) as
\begin{eqnarray}
\frac{d\left\langle{n}\right\rangle}{dt}
\!=\!-(\gamma+ W)\langle n \rangle + {\gamma n(\omega_m)+ A_ +},\label{average-phn}
\end{eqnarray}
where $W=A_--A_+$. In order to cool the MR, one needs $W>0$ (i.e.,
$A_->A_+$).

\subsection{Steady-state solution}

From Eq.~(\ref{average-phn}),
the steady-state average phonon number in the MR
is
\begin{equation}
{n}_{\rm st}=\frac{\gamma n(\omega_m)+A_+}{\gamma+W},\label{phn}
\end{equation}
%
where the term $\gamma n(\omega_m)$ in the numerator is due to the
thermal bath while $A_+$ results from the scattering processes by the
TQD. We assume that the MR is initially at equilibrium with the
thermal bath, so that the initial phonon number in the MR is
$n(\omega_m)$. In order to cool down the MR significantly, one needs a
large cooling rate $W\gg\gamma$ to overcompensate for the heating
effect of the thermal bath.  At the end of Sec.~IV, we show that this
can be achieved using typical experimental parameters.

Here we consider
$\Delta_1=\Delta_2\equiv\Delta$ so that the dark state exists. The
transition rates $A_{\pm}$ are found to be (see Appendix)
\begin{equation}
A_{\pm} = \frac{2\eta^2\Omega_1^2\Omega_2^2}{\Omega^2}
\frac{\omega_m^2\Gamma}{4[\Omega^2-\omega_m(\omega_m\pm\Delta)]^2+\omega_m^2\Gamma^2}
+\eta^2\Gamma\rho_{00}^{\rm st},\label{Apm}
\end{equation}
where $\rho_{00}^{\rm st}$ is the steady-state probability of an empty
TQD.
To cool the MR, one needs $A_->A_+$, which is
fulfilled either when $\Delta>0$ and $\Omega<\omega_m$, or when $\Delta<0$
and $\Omega>\omega_m$. Assuming also $W\gg\gamma$, the steady-state
average phonon number in the MR is approximately given by
\begin{equation}
{n}_{\rm st}\approx\frac{\gamma n(\omega_m)}{W}+n_f.\label{nst}
\end{equation}
Here $n_f\equiv{A_+}/{W}$ which gives
\begin{eqnarray}
n_f=\frac{4[\Omega^2-\omega_m(\omega_m-\Delta)]^2+\omega_m^2\Gamma^2}
{16\Delta\omega_m(\omega_m^2-\Omega^2)}.\label{nf}
\end{eqnarray}

\subsection{Optimal cooling condition}

It is easy to see that $n_f$ reaches the minimum
\begin{equation}
n_f^{\rm min}=(\frac{\Gamma}{4\Delta})^2,\label{nfmin}
\end{equation}
when the term in square brackets
in the r.h.s. of Eq.~(\ref{nf})
becomes zero, i.e.,
\begin{equation}
\Omega^2=\omega_m(\omega_m-\Delta),
\label{condition}
\end{equation}
or
\begin{equation}
\omega_m=\frac{1}{2}(\Delta+\phi).
\label{condition-2}
\end{equation}
Therefore, by properly choosing the parameters $\Omega,~\omega_m$,
and $\Delta$ so that the optimal cooling condition in
Eq.~(\ref{condition}) is fulfilled and $\Delta\gg\Gamma$, the steady-state
average phonon number in the MR can be much smaller than unity,
implying that ground-state cooling of the MR is possible.
Moreover,
the phonon number $n_f$ achievable
according to Eqs.~(\ref{nf}) and (\ref{nfmin}) is identical to the previous
results for the cooling of trapped atoms via quantum
interference~\cite{Morigi00}. However, the additional advantages of a
solid-state cooling system proposed here are that it can
be fabricated on a chip and is highly controllable. Specifically, all the relevant parameters
(i.e., the detuning $\Delta$, the tunneling rate $\Gamma$ and the
interdot coupling strengths $\Omega_1$ and $\Omega_2$) can be
controlled by tuning the gate voltages in the TQD. Thus, for a fixed
frequency $\omega_m$ of the MR, the optimal cooling condition
in Eq.~(\ref{condition}) can be conveniently fulfilled.
\begin{figure}
\includegraphics[width=3.0in,
bbllx=38,bblly=51,bburx=318,bbury=258]{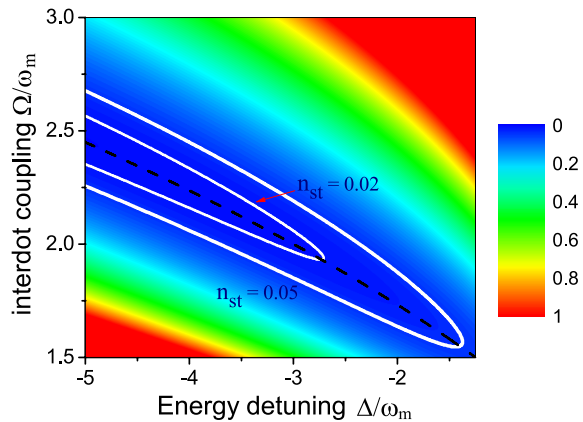}
\caption{~(color online) Contour plot of the steady-state average phonon number
${n}_{\rm st}$ in the MR as a function of the normalized driving
detuning $\Delta/\omega_m$ and the normalized interdot coupling
$\Omega/\omega_m$. The two solid curves correspond to ${n}_{\rm st}=0.05$ and $0.02$. The black dasded
line represents $\Omega^2=\omega_m(\omega_m-\Delta)$, under which
the MR can
be optimally cooled. We have chosen $\Omega_1=\Omega_2=\Omega/\sqrt{2}$ and
typical parameters $\omega_m=2\pi\times100$~MHz, $\Gamma=\omega_m$, $Q=10^5$, $\eta=0.1$,
and $n(\omega_m)=21$.}
\label{fig3}
\end{figure}

The underlying physics of the optimal cooling condition in
Eq.~(\ref{condition}) can be understood based on the eigenstate
basis of the TQD.
In the limit $\gamma\ll g\ll\omega_m$ considered here, the TQD arrives
quickly at the dark state $|-\rangle$.
The coupling between the MR and the TQD will excite the TQD to the
state $|+\rangle$ most readily when the frequency $\omega_m$ of the
MR is equal to the transition frequency $(\phi+\Delta)/2$ between
the states $|-\rangle$ and $|+\rangle$, i.e.,
$\omega_m=(\phi+\Delta)/2$. This corresponds to the transition
$|-,n\rangle\rightarrow|+,n-1\rangle$. The excited electron
subsequently tunnels to the right electrode, i.e.,
$|+,n-1\rangle\rightarrow|0,n-1\rangle$. This whole process extracts
an energy quantum from the MR. When this cycle repeats, i.e.,
$|0,n\rangle\rightarrow|-,n\rangle\rightarrow|+,n-1\rangle\rightarrow|0,n-1\rangle\rightarrow\cdots$,
the MR is cooled to the ground state. Here we emphasize that the
resonance condition for exciting the TQD from the state $|-\rangle$
to the state $|+\rangle$ via the MR is equivalent to the optimal
cooling condition in Eq.~(\ref{condition}). An electron can also
relax from the dark state $|-\rangle$ to the ground state
$|g\rangle$ by releasing energy to the MR. However, this heating
process of the MR is strongly suppressed because the frequency of
the MR is off-resonant to the transition
$|-\rangle\rightarrow|g\rangle$ in the TQD.

Figure \ref{fig3} displays a contour plot of the steady-state
average phonon number of the MR (${n}_{\rm st}$) as a function
of the effective interdot coupling $\Omega$
($=\sqrt{\Omega_1^2+\Omega_2^2}$) and the energy detuning $\Delta$.
Here we choose $\Delta<0$ and $\Omega>\omega_m$ to make sure that
$W>0$. For these typical parameters, a small ${n}_{\rm st}<0.05$
is predicted over a wide range of values on the $\Omega-\Delta$
plane. This implies that ground-state cooling of the MR should be
experimentally accessible. Furthermore, to estimate the cooling rate
$W$, we use typical experimental
parameters~\cite{TF08,WeilRevModPhys}: $\omega_m=2\pi\times100$~MHz,
$\Delta=-2\pi\times300$~MHz, and $g=2\pi\times10$~MHz. The interdot
couplings are chosen as $\Omega_1=\Omega_2\simeq2\pi\times141$~MHz
to fulfill the optimal cooling condition
$\Omega^2=\omega_m(\omega_m-\Delta)$. Using Eq.~(\ref{Apm}), one
arrives at a cooling rate $W\approx2\pi\times2$~MHz. Considering a
MR with a quality factor $Q=10^5$ (see, e.g.,
Ref.~\onlinecite{Huttel09}), one has
$\gamma=\omega_m/Q=2\pi\times1$kHz. Therefore, appreciable cooling
with $W\gg\gamma$ can be achieved. In this case, a MR can be cooled
from, e.g., an initial temperature $T=100$~mK corresponding to
$n(\omega_m)=21$ down to $T=0.8$~mK
with ${n}_{\rm st}=0.017$. 

In contrast, for sideband cooling of a
MR~\cite{Wilson04,Wilson07,Zippilli09,Ouyang09}, the
resolved-sideband cooling condition $\omega_m\gg\Gamma$ must
be followed for ground-state cooling of a MR. For a
relatively large decay rate, only MR with a very high frequency
(which becomes fragile in experiments) can be cooled. On the other
hand, for cooling via quantum interference in the TQD proposed here, the
cooling conditions $\Omega^2=\omega_m(\omega_m-\Delta)$ and
$\Delta\gg\Gamma$ do not require a high MR frequency.

\section{A scheme for verifying the cooling of the MR}

\subsection{Quantum dynamics of coupled MR-DQD system in the presence of a charge detector}

\begin{figure}
\includegraphics[width=2.0in,
bbllx=201,bblly=320,bburx=394,bbury=520]{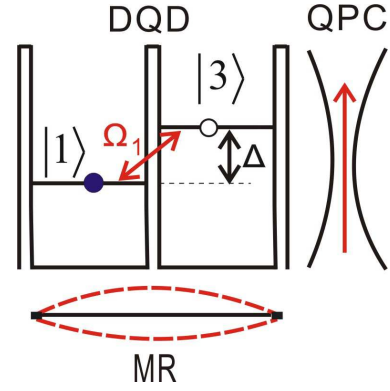}
\caption{~(color online) Schematic diagram of a MR capacitively
  coupled to a DQD which is under measurement by a nearby QPC. The energy detuning between the two dot states in the
DQD is $\Delta$ and the interdot coupling strength between them is $\Omega_1$. \label{fig4}}
\end{figure}

To verify whether the MR is cooled or not, we propose a scheme in
which the MR is coupled to a two-level system realized by a double quantum dot
(DQD).
The state of the DQD is in turn measured by a nearby charge detector
in the form of, e.g., a quantum point contact (QPC). This setup is
schematically shown in Fig.~\ref{fig4}. Experimentally, the cooling
and the verification setups are all fabricated on the same chip with
shared components. Dots 1 and 3 in the TQD from the above cooling
setup can make up the DQD after a gate voltage is applied to
decouple dot $2$ from dot $3$. Also, the QPC should be decoupled
from the rest of the system during cooling by applying a high gate
voltage.

The Hamiltonian of the system is given by
\begin{eqnarray}
H=H_R+H_{\rm DQD}+H_{\rm QPC}+H_{\rm int}+H_{\rm det}.\label{H-detect}
\end{eqnarray}
The Hamiltonian $H_R$ of
the MR and the coupling $H_{\rm int}$ between the MR and the DQD are
already given in Eqs.~(\ref{MR}) and (\ref{coupling}).
Here $H_{\rm DQD}$, $H_{\rm QPC}$ and $H_{\rm det}$ are respectively
the Hamiltonians of the DQD, the QPC and the coupling between them and
are given by
\begin{eqnarray}
H_{\rm DQD}\!&=\!&-\frac{\Delta}{2}\sigma_z+\Omega_1\sigma_x,
\nonumber\\
H_{\rm QPC}\!&=\!&\sum_{kq}\omega_{Sk}c_{Sk}^{\dagger}c_{Sk}+\omega_{Dq}c_{Dq}^{\dagger}c_{Dq},
\nonumber\\
H_{\rm det}\!&=\!&\sum_{kq}(T-\chi\sigma_z)(c_{Sk}^\dagger c_{Dq}+{\rm H.c.}),
\end{eqnarray}
where $\sigma_z=a_3^\dagger a_3-a_1^\dagger a_1$ and
$\sigma_x=a_3^\dagger a_1 + a_1^\dagger a_3$ are the Pauli matrices.
Also, $c_{ik}$ ($c_{ik}^\dagger$) is the annihilation (creation)
operator for an electron with momentum $k$ in either the source
($i=S$) or the drain ($i=D$) of the QPC. $T$  is the transition
amplitude of an isolated QPC and $\chi$ is the variation of the
transition amplitude caused by the DQD.  For simplicity, we assume
that the detunning $\Delta$ of the DQD is zero. The DQD has the
eigenstates
\begin{eqnarray}
|g\rangle=\frac{\sqrt{2}}{2}(|1\rangle-|3\rangle),~~|e\rangle=\frac{\sqrt{2}}{2}(|1\rangle+|3\rangle),
\end{eqnarray}
with $|g\rangle$ ($|e\rangle$) being the ground (excited)
state.
Rewriting the Hamiltonian in Eq. (\ref{H-detect}) on the eigenstate
basis of the DQD, we have
\begin{eqnarray}
H\!&=\!&\omega_mb^\dagger b+\Omega_1\rho_z-g\rho_x(b^\dagger+b)+H_{\rm QPC}
\nonumber\\
&&+\sum_{kq}(T-\chi\rho_x)(c_{Sk}^\dagger c_{Dq}+{\rm H.c.}),\label{H-eigenstate}
\end{eqnarray}
where $\rho_z=|e\rangle\langle e|-|g\rangle\langle g|$ and
$\rho_x=|e\rangle\langle g|+|g\rangle\langle e|$ are the Pauli
matrices.

We consider the coupled MR-DQD system in the strong dispersive
regime where the coupling strength is much smaller than the
difference between the transition frequency $2\Omega_1$ of the DQD
and that of the MR, i.e., $g\ll\delta=2\Omega_1-\omega_m$. This
regime was previously considered to study  whether
the vibration of a MR coupled to
a superconducting circuit is classical or quantum mechanical~\cite{Wei06}. In this regime, the phonon
in the MR is only virtually exchanged between the DQD and the MR.
Thus, the coupling of the DQD to the MR does not change the
occupation probability of the electron in the DQD, but only results
in phonon-number-dependent Stark shifts on energy levels of the DQD.
Moreover, the Stark shifts can be detected by measuring the
full-frequency current spectrum of the QPC.

Applying both a rotating-wave approximation and a canonical
transformation $U'=e^{s'}$ with
\begin{equation}
s'=\eta(\rho_-b^\dagger-\rho_+b),~~\eta=g/\delta,
\end{equation}
to the Hamiltonian $H$, one obtains [up to $\mathcal{O}(\eta^2)$]
\begin{eqnarray}
H\!&\approx&\!\omega_mb^\dagger b+\frac{2\Omega_1+{g^2}(2b^\dagger b+1)/{\delta}}{2}\rho_z
\nonumber\\
&&+\frac{g^2}{2\delta}(|g\rangle\langle g|+|e\rangle\langle e|)+H_{\rm QPC}
\nonumber\\
&&+\sum_{kq}(T-\chi\rho_x)(c_{Sk}^\dagger c_{Dq}+{\rm H.c.}).\label{H-approx}
\end{eqnarray}
From Eq.~(\ref{H-approx}), after taking the trace over the degrees
of freedom of the QPC, one obtains the following master equation for the
reduced density matrix elements of the coupled MR-DQD
system~\cite{Backaction}:
\begin{eqnarray}
\dot{\rho}_{gn,gn}\!&=\!&-\gamma_+\rho_{gn,gn}+(\gamma_-+\gamma_d)\rho_{en,en},
\nonumber\\
\dot{\rho}_{en,en}\!&=\!&\gamma_+\rho_{gn,gn}-(\gamma_-+\gamma_d)\rho_{en,en},
\nonumber\\
\dot{\rho}_{gn,en}\!&=\!&-i\delta_n\rho_{gn,en}-(\gamma_1+\frac{\gamma_d}{2})\rho_{gn,en},\label{ME-det}
\end{eqnarray}
where $\delta_n=2\Omega_1+{g^2}(2n+1)/\delta$ and $\gamma_1=2\pi
g_sg_d\chi^2eV_d$ with $g_s$ ($g_d$) being the density of states for
electrons in the source (drain) of the QPC and $V_d$ the bias
voltage across the QPC. Here $\gamma_{\pm}=\gamma_1(1\mp\lambda_n)$,
with $\lambda_n=\delta_n/eV_d$, are the QPC-induced excitation and
relaxation rates between the ground state and the excited state of
the DQD. Also, $\gamma_d$ is the relaxation rate resulting from the
coupling of the DQD to the thermal bath. Since the dissipation rate
of the MR is much smaller than that of the DQD, dissipation of the
MR is neglected. In Eq. (\ref{ME-det}), the reduced density matrix
element $\rho_{in,in}$ ($i=g,e$) gives the occupation probability of
the state $|i,n\rangle$ of the coupled MR-DQD system ,while
$\rho_{in,jn}$ ($i\neq j$) describes the coherence between the
states $|i,n\rangle$ and $|j,n\rangle$. The equations of motion for
other elements, e.g., $\rho_{in,jn'}$ ($n\neq n'$), which are
decoupled from those considered here, are not shown.
Using Eq.~(\ref{ME-det}) and the normalization condition
$p_n=\rho_{gn,gn}+\rho_{en,en}$, one finds
\begin{eqnarray}
\rho_{gn,gn}(t)\!&\!=\!&\!\frac{(\gamma_-+\gamma_d)p_n}{2\gamma_0}
\nonumber\\
&&-\big[\frac{(\gamma_-+\gamma_d)p_n}{2\gamma_0}-\rho_{gn,gn}(0)\big]e^{-2\gamma_0t},
\nonumber\\
\rho_{en,en}(t)\!&\!=\!&\!\frac{\gamma_+}{2\gamma_0}p_n
-\big[\frac{\gamma_+}{2\gamma_0}p_n-\rho_{en,en}(0)\big]e^{-2\gamma_0t},
\nonumber\\
\rho_{gn,en}(t)\!&\!=\!&\!\rho_{gn,en}(0)e^{-i(\delta_n-\gamma_0)t},\label{analyticalexpression}
\end{eqnarray}
where $\gamma_0=\gamma_1+\gamma_d/2$ and $p_n$ is the probability
that the MR is at state $|n\rangle$.

\subsection{Current spectrum of the charge detector}

The dc current through the QPC is related to the electron occupation
probability in the DQD and is given
by~\cite{GurvitzPRB97}
\begin{equation}
I(t)\!=eD\rho_{11}+eD'\rho_{33}=\frac{e}{2}(D+D')+\frac{e}{2}(D'-D)\langle \sigma_z\rangle,
\end{equation}
where
\begin{eqnarray}
D\!&\!=\!&\!2\pi g_sg_d(T-\chi)^2V_d,\nonumber\\
D'\!&\!=\!&\!2\pi g_sg_d(T+\chi)^2V_d,
\end{eqnarray}
are the respective rates of electron tunneling through the QPC when dot
$3$ is respectively occupied or empty~\cite{GurvitzPRB97}. Therefore, one can
define the current operator as
\begin{eqnarray}
I(t)=I_0+I_1\sigma_z(t)=I_0+I_1\rho_x(t),\label{currentoperator}
\end{eqnarray}
with $I_{0,1}=e(D\pm D')/2$. According to the Wiener-Khintchine
theorem, the power spectrum of the current through the QPC
is~\cite{Scully}
\begin{equation}
S(\omega)\!=\!{\rm Re}\!\!\int\limits_0^{\infty} e^{i\omega\tau}d\tau[\langle I(t)I(t+\tau)\rangle
-\langle I(t+\tau\rangle\langle I(t)\rangle].
\label{currentspectrum}
\end{equation}
Substituting Eqs.~(\ref{analyticalexpression}) and
(\ref{currentoperator}) into Eq.~(\ref{currentspectrum}), we get
\begin{eqnarray}
S(\omega)/S_0\!&\!=\!&\!1+\frac{2\gamma_1\gamma_2}{\gamma_1+\gamma_2}
\sum_np_n(1-\kappa p_n)\frac{\gamma_0}{\gamma_0^2+(\delta_n-\omega)^2}
\nonumber\\
&&-\frac{2\gamma_1\gamma_2}{\gamma_1+\gamma_2}\sum_np_n(1+\kappa p_n)
\frac{\gamma_0}{\gamma_0^2+(\delta_n+\omega)^2},
\nonumber\\&&
\label{spectrum}
\end{eqnarray}
where $\gamma_2=2\pi g_sg_dT^2V_d$,
$\kappa=(\gamma_+-\gamma_--\gamma_d)/2\gamma_0$, and $S_0=2eI_0$ is
the current-noise background. From Eq.~(\ref{spectrum}), one sees
that the current spectrum of the QPC consists of peaks at resonance
points $\omega=\pm\delta_n$. These peaks have width $\gamma_0$ and
heights increasing with the probability $p_n$. For instance, the
peaks at the resonance point
\begin{equation}
\delta_n=2\Omega_1+\frac{g^2(2n+1)}{\delta},
\end{equation}
is shifted by ${g^2}(2n+1)/\delta$ from $2\Omega_1$. Thus, from this
peak shift in the current spectrum, one can readout the
phonon-number state of the MR.
\begin{figure}
\includegraphics[width=3.0in,
bbllx=16,bblly=48,bburx=300,bbury=258]{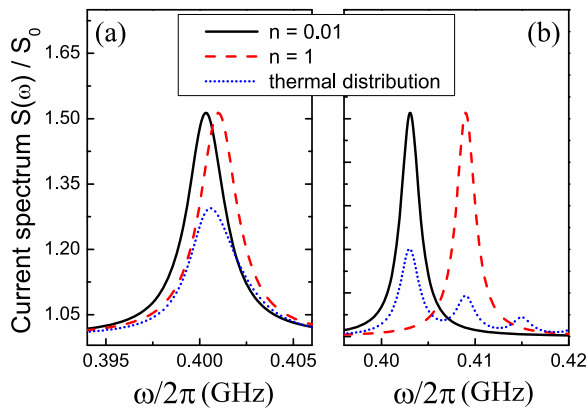}
\caption{~(color online) Power spectrum of the current through the QPC
when the phonon number in the MR are respectively $n=0.01$ (black solid line),
$n=1$ (red dashed line),
or given by the thermal distribution (blue dotted line), i.e.,
$p_n={n}_{\rm st}^n/{(1+{n}_{\rm st})^{n+1}}$ with ${n}_{\rm st}=1$.
The coupling strength between the MR and the DQD is
$g=0.1~\omega_m$ (a) and $g=0.3~\omega_m$ (b). The other parameters are
$\omega_m=100~$MHz, $\Omega_1=2\omega_m$, $\gamma_2=0.01\omega_m$,
$\gamma_1=0.2\gamma_2$, and $\gamma_d=2\gamma_2$.  \label{fig5}}
\end{figure}

Figure 4 plots the current spectrum of the QPC with two different
coupling strengths between the MR and the DQD. Results for three
cases in which the MR is respectively in its ground state
($n=0.01\ll1$), the first-excited state ($n=1.0$) or thermalized
with an average phonon number ${n}_{\rm st}=1.0$ are plotted. Each
resonance peak in the current spectrum corresponds to a
phonon-number state of the MR. For the thermally distributed case,
the current spectrum shows several peaks, where the relative area
under each peak gives the probability of the corresponding
phonon-number state. As shown in Fig.~\ref{fig4}(a), the distance
between two adjacent peaks is smaller than the intrinsic peak width
in the weak dispersive regime, i.e., $2g^2/\delta<\gamma_0$, and
hence the measured spectrum shows an ensemble. In this case, the
phonon-number state of the MR cannot be measured. In the strong
limit ($2g^2/\delta>\gamma_0$), however, the ensemble can be
individually resolved [Fig.~\ref{fig4}(b)], which allows us to
detect the phonon number and also to verify the cooling result of
the MR. Indeed, a relatively strong coupling between a MR and a
quantum dot has been recently demonstrated~\cite{Huttel09}. The
strong dispersive regime is thus achievable and one can apply the
proposed coupled MR-DQD system to verify the cooling of the MR via
measuring the current spectrum of a nearby charge detector (e.g.,
QPC).

\section{Discussion and Conclusion}

Our proposal on ground state cooling of the MR requires that the TQD
is able to evolve into the dark state. However, the dephasing of the
TQD due to coupling to other degrees of freedom in the environment
can project the TQD into one of the three localized states
$|1\rangle$, $|2\rangle$, and $|3\rangle$ and drive the system away
from the dark state~\cite{Michaelis06}. However, the dephasing
between the two localized states $|1\rangle$ and $|2\rangle$ depends
on the coupling strength and the energy detunning between dots $1$
and $2$~\cite{BrandesPRL00}. Here in our system no direct coupling
exists between the two localized states $|1\rangle$ and $|2\rangle$,
and thus the dephasing almost has negligible effects on the cooling
efficiency of the MR.

In summary, we have studied the cooling of a MR by capacitive
coupling to a TQD. We show that when the two lower-energy localized states
become degenerate, the TQD will be trapped in a dark state which is
decoupled from the excited state in the absence of the MR.
With the MR in resonance with the transition between the dark state
and the excited eigenstate in the TQD, we have shown that the MR can
be cooled to its ground state in the {\it non-resolved sideband
cooling regime}. Moreover, we have proposed a coupled MR-DQD system
in the strong dispersive regime for verifying the cooling result of
the MR. In this regime, the coupling between the MR and the DQD
induces a MR-phonon-number dependent shift of the transition
frequency of the DQD. Thus the phonon-number state which
characterizes the cooling result of the MR can be detected by
measuring the shifts of the resonance peaks in the current spectrum
of a nearby charge detector.

\begin{acknowledgments}
This work is supported by the National Basic Research Program of
China Grant Nos. 2009CB929300 and 2006CB921205, the National Natural
Science Foundation of China Grant Nos. 10534060 and 10625416, and
the Research Grant Council of Hong Kong SAR project No. 500908.
\end{acknowledgments}

\appendix
\section{Master equation for the reduced density matrix of the MR}

In this appendix, we derive the master equation [Eq. (\ref{ME-MR})]
for the reduced density matrix of the MR from the master equation
[Eq. (\ref{ME})] of the coupled MR-TQD system by eliminating the
degrees of freedom of the TQD. In general, the dissipation rate of
the MR is much smaller than the decay rate of the TQD, i.e.,
$[n(\omega_m)+1]\gamma\ll\Gamma$. The TQD hence attains its
steady-state quickly and its perturbation to the MR can be regarded
as part of the environment~\cite{Wilson07,Cirac92}. Up to the second
order in $\eta$, Eq.~(\ref{ME}) can be rewritten as
\begin{eqnarray}
\frac{d\rho}{dt}\!=\!\mathcal{L}\rho=[\mathcal{L}_0+\mathcal{L}_1+\mathcal{L}_2]\rho,
\end{eqnarray}
where
\begin{eqnarray}
\mathcal{L}_0{\rho}\!&=\!&-i\omega_m[b^\dagger b,\,\rho]-i[H_{\rm TQD},\rho]
\nonumber\\
&&+\Gamma_1\mathcal{D}[a_1^\dagger]\rho+\Gamma_2\mathcal{D}[a_2^\dagger]\rho
+\Gamma_3\mathcal{D}[a_3]\rho,\label{L0}\\
\mathcal{L}_1\rho\!&=\!&-i[V(b^\dagger-b),\rho],\label{L1}\\
\mathcal{L}_2\rho\!&=\!&\eta^2\Gamma_1(\mathcal{D}[a_1^\dagger b^\dagger]\rho+\mathcal{D}[a_1^\dagger b]\rho)
\nonumber\\
&&+\eta^2\Gamma_3(\mathcal{D}[a_3b^\dagger]\rho+\mathcal{D}[a_3b]\rho)+\mathcal{L}_{\rm D}\rho,\label{L2}
\end{eqnarray}
are respectively the Liouvillians to zeroth, first, and second orders
in $\eta$. At zeroth order in $\eta$, the quantum dynamics of the
whole system is described by
\begin{equation}
\dot{\rho}(t)=\mathcal{L}_0\rho(t).\label{A-1}
\end{equation}
The MR and the TQD are decoupled.
Since the TQD is at its steady state most of the time, one has
$\rho(t)=\rho_d^{\rm st}\otimes{\rm Tr}_{d}\{\rho\}$ with $\rho_d^{\rm
st}$
denoting the reduced density matrix of the TQD at steady-state and
${\rm Tr}_{d}\{\cdots\}$ the trace over the TQD's degrees of
freedom. Equation~(\ref{A-1}) has an infinite number of steady-state
solutions.  These solutions can be expanded in the basis of the
eigenvectors $\rho_d^{\rm st}\otimes|n\rangle\langle n'|$ of the
Liouville operator $\mathcal{L}_0$ with eigenvalues
$\lambda_{nn'}=-i(n-n')\omega_m$, i.e.,
%
$\mathcal{L}_0\rho_{nn'}
=\lambda_{nn'}\rho_{nn'}$~\cite{Morigi00}.
Here, $|n\rangle$ ($n=0, 1, 2, \cdots$) denotes the $n$th state of the MR  and
$(n-n')\omega_m$ represents the energy difference between the states
$|n\rangle$
and $|n'\rangle$. For $\eta\neq0$, these states
with different $n$ are weakly coupled by the perturbative terms
$\mathcal{L}_1$ and $\mathcal{L}_2$. To obtain the quantum dynamics of
the MR, we project the system onto the
subspace with a zero eigenvalue ($n=n'$) of $\mathcal{L}_0$. The
projection operator $\mathcal{P}$ is defined by
\begin{equation}
\mathcal{L}_0\mathcal{P}\rho=0.
\end{equation}
Noting
$g\ll\omega_m$ (i.e., $\eta\ll1$), a second order perturbation
expansion gives the following closed equation for $\mathcal{P}\rho$
~\cite{Ouyang09}
\begin{eqnarray}
\mathcal{P}\dot{\rho}(t)=\mathcal{P}\mathcal{L}_2\mathcal{P}\rho(t)
+\int\limits_0^\infty d\tau\mathcal{P}\mathcal{L}_1e^{\mathcal{L}_0\tau}\mathcal{L}_1\mathcal{P}\rho(t).\label{APP}
\end{eqnarray}
Substituting Eqs.~(\ref{L1}) and (\ref{L2}) into Eq.~(\ref{APP}) and
taking the trace over the TQD degrees of freedom, the
first term in Eq.~(\ref{APP}) becomes~\cite{Cirac92}
\begin{eqnarray}
{\rm Tr}_d\{\mathcal{P}\mathcal{L}_2\mathcal{P}\rho(t)\}
\!&\!=\!&\!\frac{1}{2}[\gamma n(\omega_m)+\eta^2\Gamma_1\rho_{00}^{\rm st}]\mathcal{D}[b]\mu
\nonumber\\
&&+\frac{1}{2}\big\{\gamma [n(\omega_m)+1]+\eta^2\Gamma_1\rho_{00}^{\rm st}\big\}\mathcal{D}[b^\dagger]\mu,
\nonumber\\
&&\label{firstterm}
\end{eqnarray}
and the second term gives
\begin{eqnarray}
&&{\rm Tr}_d\bigg\{\int\limits_0^\infty d\tau
\mathcal{P}\mathcal{L}_1e^{\mathcal{L}_0\tau}\mathcal{L}_1\mathcal{P}\rho(t)\bigg\}
\nonumber\\
\!&\!=\!&\!-i(\omega_m+\delta_m)[b^\dagger b,\mu],
\nonumber\\
&&+{\rm Re}[G(i\omega_m)]\mathcal{D}[b]\mu+{\rm Re}[G(-i\omega_m)]\mathcal{D}[b^\dagger]\mu,
\label{secondterm}
\end{eqnarray}
where $\mu={\rm Tr}_d\{\mathcal{P}\rho\}$ is the reduced density
matrix of the MR and $\rho_{00}^{\rm st}$ is the probability of an
empty TQD at the steady state. Here, we have defined
\begin{equation}
\delta_m={\rm Im}[G(i\omega_m)+G(-i\omega_m)].
\end{equation}
Thus, from Eqs.~(\ref{APP}), (\ref{firstterm}) and
(\ref{secondterm}), one has
\begin{eqnarray}
\dot{\mu}&=&-i(\omega_m+\delta_m)[b^\dagger b,\mu]
+\frac{1}{2}[\gamma n(\omega_m)+A_+]\mathcal{D}[b]\mu
\nonumber\\
&&
+\frac{1}{2}\big\{\gamma [n(\omega_m)+1]+A_-\big\}\mathcal{D}[b^\dagger]\mu,\label{A-ME-MR}
\end{eqnarray}
where
\begin{equation}
A_{\pm}=2\,{\rm Re}[G(\pm i\omega_m)]+\eta^2\Gamma_1\rho_{00}^{\rm st}.
\label{A}
\end{equation}
Eq.~(\ref{A-ME-MR}) is simply the master equation (\ref{ME-MR}) for
the reduced density matrix of the MR derived in Sec.~IV. In
Eq.~(\ref{secondterm}), the correlation function $G(s)$ is given by
\begin{eqnarray}
G(s)\!&\!=\!&\!-{\rm Tr}_d\int\limits_0^\infty dt V(0)e^{\mathcal{L}_{\rm TQD}t}V(0)e^{st}
\nonumber\\
\!&\!=\!&\!-\int\limits_0^\infty d\tau \langle V(t)V(0)\rangle e^{st}
\!=\!-\langle\widetilde{V}(s)V(0)\rangle,
\nonumber\\&&~~~
\label{correlationfunction}
\end{eqnarray}
where $s=i\omega_m$. Also, $\widetilde{V}(s)$ is the Laplace transform
of $V(t)$ and the Liouvillians $\mathcal{L}_{\rm TQD}$ is given in
Eq.~(\ref{ME-TQD}).

To determine the correlation function $G(\pm s)$, one first calculate the Laplace transform $\widetilde{V}(s)$ of the
interaction term $V(t)$. For convenience, we introduce the vector
operator $\hat{\sigma}$ for the TQD whose components are defined as
\begin{eqnarray}
&&\hat{\sigma}_1=|1\rangle\langle1|,~\hat{\sigma}_2=|2\rangle\langle2|,~\hat{\sigma}_3=|3\rangle\langle3|,
\nonumber\\
&&\hat{\sigma}_4=|1\rangle\langle2|,~\hat{\sigma}_5=|2\rangle\langle1|,~\hat{\sigma}_6=|1\rangle\langle3|,
\nonumber\\
&&\hat{\sigma}_7=|3\rangle\langle1|,~\hat{\sigma}_8=|2\rangle\langle3|,~\hat{\sigma}_9=|3\rangle\langle2|,
\end{eqnarray}
where the average value of each component is
$\langle\hat{\sigma}_i\rangle={\rm Tr}\{\hat{\sigma}_i\rho_d\}$.
Using this notation, one has
\begin{equation}
V=2\eta\Omega_1(\hat{\sigma}_7-\hat{\sigma}_6)+\eta\Omega_2(\hat{\sigma}_9-\hat{\sigma}_8).
\label{v}
\end{equation}
and thus
\begin{eqnarray}
G(s)
\!&\!=\!&\!-2\eta\Omega_1[S_7(s)-S_6(s)]-\eta\Omega_2[S_9(s)-S_8(s)],
\label{G}\nonumber\\&&
\end{eqnarray}
where $S_i(s)=\langle \widetilde{\sigma}_i(s)V(0)\rangle$ with
$\langle \widetilde{\sigma}_i(s)\rangle$ being the Laplace transform
of $\langle\hat{\sigma}_i(t)\rangle$. From Eq.~(\ref{ME-TQD}), we find
that $\langle\hat{\sigma}_i(t)\rangle$ obeys the equation of
motion:
\begin{equation}
\frac{d{\langle\hat{\sigma}(t)\rangle}}{dt}=M{\langle\hat{\sigma}(t)\rangle}+B,\label{EOM-TQD}
\end{equation}
where
\begin{widetext}
\begin{eqnarray}
\begin{array}{l}
 M\!=\!\left( {\begin{array}{*{20}c}
   -\Gamma_1 & -\Gamma_1 & -\Gamma_1 & 0 & 0 & -i\Omega_1 & i\Omega_1 & 0 & 0 \\
   -\Gamma_2 & -\Gamma_2 & -\Gamma_2 & 0 & 0 & 0 & 0 & -i\Omega_2 & i\Omega_2 \\
   0 & 0 & -\Gamma_3 & 0 & 0 & i\Omega_1 & -i\Omega_1 & i\Omega_2 & -i\Omega_2 \\
   0 & 0 & 0 & -i\Delta_d & 0 & -i\Omega_2 & 0 & 0 & i\Omega_1 \\
   0 & 0 & 0 & 0 & i\Delta_d & 0 & i\Omega_2 & -i\Omega_1 & 0 \\
   -i\Omega_1 & 0 & i\Omega_1 & -i\Omega_2 & 0 & \lambda_1 & 0 & 0 & 0 \\
   i\Omega_1 & 0 & -i\Omega_1 & 0 & i\Omega_2 & 0 & \lambda_1^* & 0 & 0 \\
   0 & -i\Omega_2 & i\Omega_2 & 0 & -i\Omega_1 & 0 & 0 & \lambda_2 & 0 \\
   0 & i\Omega_2 & -i\Omega_2 & i\Omega_1 & 0 & 0 & 0 & 0 & \lambda_2^* \\
\end{array}} \right), \\
  \end{array}\label{MatrixM}
  \nonumber\\
\end{eqnarray}
\end{widetext}
and $B=(\Gamma_1,\Gamma_2,0,0,0,0,0,0,0)^T$. Here we have defined
$\Delta_d\equiv\Delta_1-\Delta_2$,
$\lambda_1\equiv-(i\Delta_1+\frac{1}{2}\Gamma_3)$, and
$\lambda_2\equiv-(i\Delta_2+\frac{1}{2}\Gamma_3)$. From
Eq.~(\ref{EOM-TQD}), the steady-state solution of the vector
$\langle\hat{\sigma}\rangle$ is calculated as
\begin{equation}
\langle\hat{\sigma}^{\rm st}\rangle=M^{-1}B.\label{steadysolution}
\end{equation}
Applying the Laplace transform to Eq.~(\ref{EOM-TQD}), one obtains
\begin{equation}
s\langle\widetilde{{\sigma}}(s)\rangle-\langle\hat{\sigma}(0)\rangle
=M\langle\widetilde{{\sigma}}(s)\rangle+\frac{B}{s}.
\end{equation}
Moreover, the $i$th component of the vector $\langle\widetilde{\sigma}(s)\rangle$ is given by
\begin{equation}
\langle\widetilde{{\sigma}}_i(s)\rangle=\sum_{k=1}^9L_{ik}[\langle\hat{\sigma}_k(0)\rangle+\frac{B_k}{s}].
\label{sigma-s}
\end{equation}
Here the matrix $L$ is defined as $L=(sI-M)^{-1}$ where $I$ denotes
the identity matrix. Assuming that the TQD has already attained its
steady-state at initial time $t=0$, i.e.,
$\langle\hat{\sigma}(0)\rangle=\langle\hat{\sigma}^{\rm st}\rangle$,
using Eqs.~(\ref{v}), (\ref{G}), (\ref{steadysolution}),
(\ref{sigma-s}) and the quantum regression theorem~\cite{Scully},
one can obtain the correlation function $G(s)$ and the scattering
rates $A_{\pm}$ as given in Eq.~(\ref{A}).


\end{document}